\documentclass[sigconf, screen]{acmart}
\AtBeginDocument{%
  }

\setcopyright{acmlicensed}
\copyrightyear{2025}
\acmYear{2025}
\acmDOI{XXXXXXX.XXXXXXX}
\acmConference[Conference acronym 'XX]{Make sure to enter the correct
  conference title from your rights confirmation email}{June 03--05,
  2025}{Woodstock, NY}
\acmISBN{978-1-4503-XXXX-X/2025/02}

\begin{document}

\title{Interacting with \textit{Thoughtful} AI}


\author{Xingyu Bruce Liu}
\affiliation{%
  \institution{University of California, Los Angeles}
  \city{Los Angeles}
  \state{CA}
  \country{USA}}
 \email{xingyuliu@ucla.edu}

 \author{Haijun Xia}
\affiliation{%
  \institution{University of California, San Diego}
  \city{La Jolla}
  \state{CA}
  \country{USA}}
 \email{haijunxia@ucsd.edu}

\author{Xiang `Anthony' Chen}
\affiliation{%
  \institution{University of California, Los Angeles}
  \city{Los Angeles}
  \state{California}
  \country{USA}}
\email{xac@ucla.edu}

\renewcommand{\shortauthors}{Liu et al.}

\begin{abstract}

We envision the concept of \textit{Thoughtful AI}, a new human–AI interaction paradigm in which the AI behaves as a continuously thinking entity.
Unlike conventional AI systems that operate on a turn-based, input-output model, Thoughtful AI autonomously generates, iterates, and communicates its intermediate thought process throughout an interaction.
In this position paper, we argue that this thoughtfulness unlocks new possibilities for human-AI interaction by enabling proactive AI behavior, facilitating continuous alignment with users, and fostering more dynamic interaction experiences.
We outline the conceptual foundations of Thoughtful AI, illustrate its potential through example systems, and envision how this paradigm can transform human-AI interaction in the future.

\end{abstract}


\begin{teaserfigure}
\end{teaserfigure}


\maketitle

\section{Introduction}

\begin{figure}
    \centering
  \includegraphics[width=\linewidth]{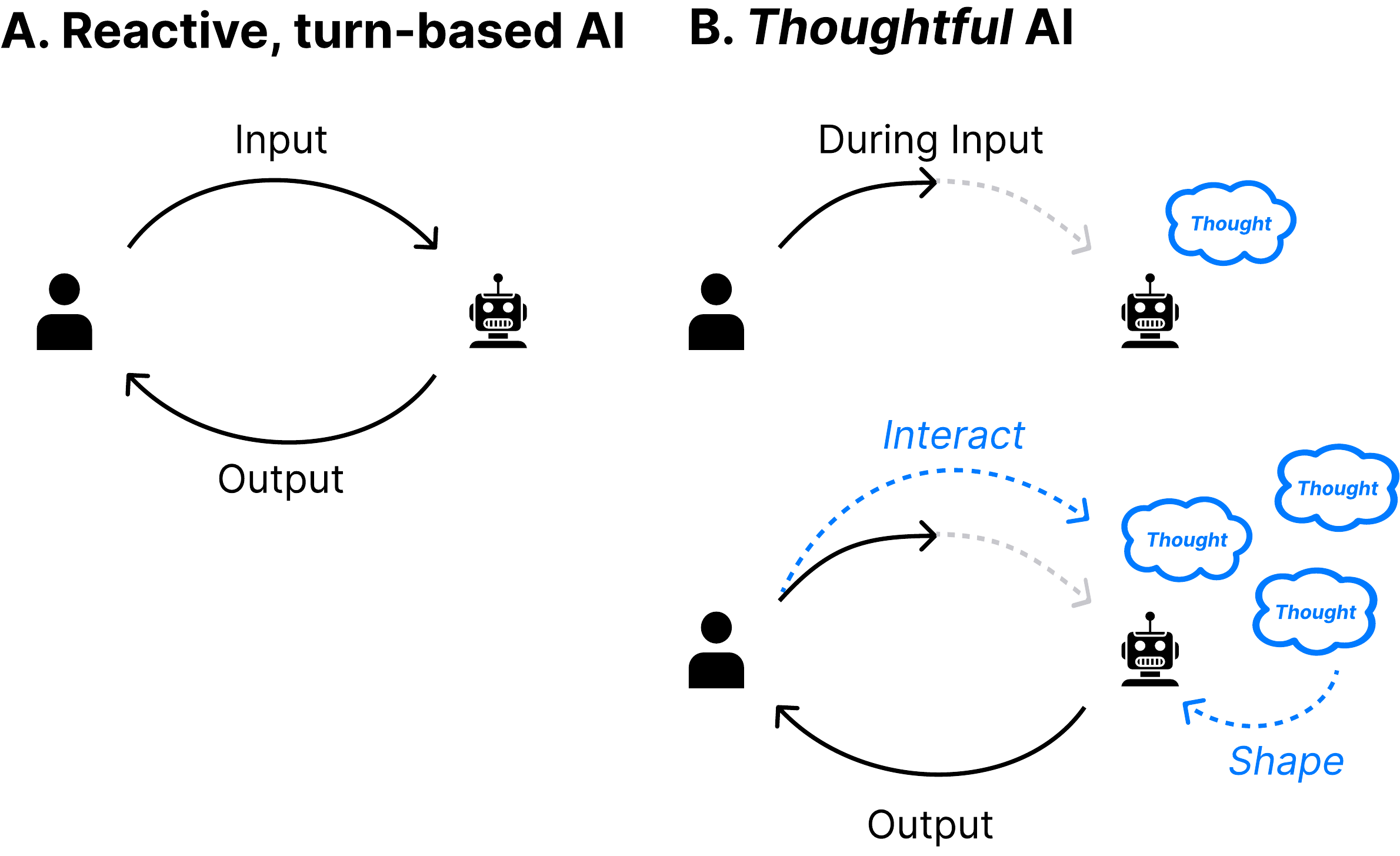}
    \caption{Rethinking human-AI interaction paradigms:
(A) Traditional AI is reactive, responding only when prompted.
(B) Thoughtful AI thinks continuously, proactively generating, iterating, and allowing users to interact with its thoughts.
}
    \label{fig:teaser}
\end{figure}

In 1950, Alan Turing famously asked, \textit{“Can machines think?”}~\cite{Turing1950ComputingMA} This question has inspired decades of research in artificial intelligence (AI), with efforts ranging from symbolic reasoning and expert systems to today’s large language models (LLMs).
Early philosophical and computational models of cognition viewed thought as a defining characteristic of intelligence—whether in humans or machines~\cite{descartes1996discourse, newell1972human, miller2001integrative, raichle2001default}. 
Recent LLMs have reignited discussions about what it means for an AI to ``think''.
Techniques such as Chain-of-Thought prompting~\cite{wei2022chain} demonstrate that generating intermediate reasoning steps can significantly enhance model performance on complex tasks. Building on this foundation, models like OpenAI o1~\cite{jaech2024openai}, Gemini Flash Thinking~\cite{deepmind2025geminiflashthinking}, and DeepSeek R1~\cite{guo2025deepseek} integrate reasoning into their training processes. These developments mark substantial progress in AI’s ability to problem-solve.

However, despite these advancements, current human-AI interaction paradigms remain fundamentally constrained. Existing systems still operate within a turn-based, input-output framework: Users issue a prompt, the AI generates a single response, and the cycle repeats. 
This limitation becomes evident in everyday AI interactions. Imagine a user asking ChatGPT: ``Help me plan a surprise birthday party for my friend.'' 
ChatGPT would generate a static response---perhaps listing general steps like choosing a venue, sending invitations, and selecting a menu. 
At this point, the AI has already stopped processing and just passively waits for the user’s next input. 
If the user later realizes they need a budget-friendly vegan restaurant for six people in New York, they must start a new turn, manually refining their query.

In contrast, consider instead how humans naturally collaborate during a brainstorming session. Each participant holds an internal train of thought that continuously updates in the background, even while someone else is speaking. In parallel, they may externalize immature, intermediate thoughts by sketching or annotating on a whiteboard, inviting immediate refinements and comments from others. This thinking process does not pause in discrete cycles: conversation flows as people update, erase, and reorganize ideas in real time.

In this position paper, we envision the concept of \textit{Thoughtful AI}, a new interaction paradigm in which AI functions as a continuously thinking entity. 
Unlike conventional AI that passively waits for user prompts and responds in discrete turns, Thoughtful AI autonomously generates, iterates, and selectively communicates its evolving thoughts throughout an interaction.

At the core of Thoughtful AI is to view AI's \textit{thought} as a first-class citizen in human-AI interaction: it is not merely hidden backend computation steps but a fundamental modality that contribute to AI's capabilities to interact with humans. 
We identify four key traits that define this paradigm:
\begin{enumerate} 
    \item It provides an \textit{intermediate medium}, enabling users to observe and interact with AI’s intermediate thoughts rather than just its final outputs. 
    \item It enables a \textit{full-duplex process}, where AI and users exchange thoughts fluidly rather than in rigid, turn-based exchanges. 
    \item It serves as a \textit{intrinsic driver}, allowing AI to initiate interactions rather than merely responding to queries. 
    \item It establishes a \textit{shared cognitive space}, where AI and users build upon each other’s thoughts in a collaborative, dynamic process.
\end{enumerate}

In the sections that follow, we first outline the conceptual foundations of Thoughtful AI.
We then provide concrete examples that illustrate how a continuously thinking AI can bring new possibilities to human-AI interaction, offering benefits like enabling proactive AI behavior, facilitating continuous cognitive alignment with users, and fostering more dynamic interaction experience.
We discuss the broader implications of this paradigm shift, and explore how ``thinking’’---long considered a uniquely human trait---can transform our ways to interact with machines in the future.

\section{What makes AI ``Thoughtful''?}

Discussions around ``thought'' in LLM research often focus on a chain-of-thought approach, where a model generates intermediate reasoning steps to improve performance on reasoning tasks. From an HCI perspective, however, \textit{thought} can be understood more broadly. We define \textit{Thoughtfulness} as:

\newtheorem{thoughtfuldefinition}{Definition}
\begin{definition}
\textbf{Thoughtfulness} refers to a system's ability to continuously generate, develop, and selectively communicate its \textit{intermediate processes and responses} over the course of an interaction. 
\end{definition}

In contrast to LLM-based definitions, our conception of Thoughtful AI emphasizes how these ongoing processes shape the system’s behavior and its capacity to interact with human. Thoughts can emerge at any point in an interaction, triggered by external stimuli or intrinsic reflection, may be expressed or remain internal, and they can take various forms, from abstract keywords to visual or auditory representations.

To better understand how Thoughtful AI differs from conventional systems, we examine four key traits of this interaction paradigm (\autoref{tab:02-traits}), comparing them with the turn-based model to highlight the key distinctions.

\subsection{Traits of Thoughtful AI}


\begin{table}
  \resizebox{\columnwidth}{!}{%
    \begin{tabular}{@{}lll@{}}
      \toprule
      \textbf{Trait} & \textbf{Current AI} & \textbf{Thoughtful AI} \\ 
      \midrule
      \textit{Intermediate Medium} & 
        \begin{tabular}[c]{@{}l@{}}
          Reveals only\\
          final outputs
        \end{tabular} & 
        \begin{tabular}[c]{@{}l@{}}
          Surfaces intermediate\\
          thoughts in real time
        \end{tabular} \\
      \addlinespace
      \textit{Full-Duplex Process} & 
        \begin{tabular}[c]{@{}l@{}}
          Waits for user\\
          prompts
        \end{tabular} & 
        \begin{tabular}[c]{@{}l@{}}
          Continuously thinks in\\
          parallel with user activity
        \end{tabular} \\
      \addlinespace
      \textit{Intrinsic Driver} & 
        \begin{tabular}[c]{@{}l@{}}
          Responds only\\
          when asked
        \end{tabular} & 
        \begin{tabular}[c]{@{}l@{}}
          Self-initiates actions\\
          based on thoughts
        \end{tabular} \\
      \addlinespace
      \textit{Shared Cognitive Space} & 
        \begin{tabular}[c]{@{}l@{}}
          Turn-based\\
          exchanges
        \end{tabular} & 
        \begin{tabular}[c]{@{}l@{}}
          Builds on collaborative\\
          fragmented ideas
        \end{tabular} \\
      \bottomrule
    \end{tabular}%
  }
  \caption{Comparison of Thoughtful AI and Current AI across key traits.}
  \label{tab:02-traits}
\end{table}

\subsubsection{Intermediate Medium}
A fundamental distinction between a \textit{thought} and a \textit{response} is that a thought is \textit{in-progress}, potentially fragmented, and not necessarily intended for final output. In Thoughtful AI, the system’s processing, whether it involves brainstorming ideas, weighing pros and cons, or exploring parallel strategies, can be treated as an \textit{intermediate medium} that the AI may optionally share.

In most existing AI systems, users see only a final answer. This one-shot approach is limiting: once the user receives an answer, any clarifications or follow-up requests often require multiple back-and-forth queries. If the AI’s thinking was flawed or incomplete, the user has no window into how or why the system arrived at its conclusion.

\subsubsection{Full-duplex Process}
Human conversation does not pause for one party to ``finish thinking'', nor do humans stop thinking when others start to speak. Similarly, Thoughtful AI enables a \textit{full-duplex process} where thinking is continuous and can occur any point in the interaction, rather than locked into turn-based cycles. AI thoughts may be triggered by user inputs or by the system’s own internal reflections; they can also evolve without any external stimuli, for example, during periods of user silence.

Conventional AI interactions are typically half-duplex. After producing an answer, the AI stops listening or processing until another user query arrives. It remains idle, unaware of changing contexts and does not develop anything internally during that downtime.

\subsubsection{Intrinsic Driver}
One of the key aspects of human intelligence is that thought is not merely reactive: it also serves as an \textit{intrinsic driver} of action. Similarly, in Thoughtful AI, thinking can serve as a mechanism that enables the system to self-initiate its interactions and be \textit{proactive}.
Rather than being solely triggered by user input, Thoughtful AI can independently generate ideas, monitor evolving contexts, and identify opportunities to intervene or contribute.

Traditional turn-based systems are \emph{reactive}: they respond only once the user issues a query. Even so-called ``proactive'' assistants typically rely on predefined triggers or simple rule-based heuristics. In contrast, a Thoughtful AI continuously \emph{thinks} in the background, allowing it to model its intrinsic motivation to take actions based on the intermediate thoughts. Its continuous thought process is the primary engine for proactive behavior.

\subsubsection{Shared Cognitive Space}

Finally, a \textit{shared cognitive space} could emerge when AI and user co-exist in an ongoing thought process. Rather than the AI presenting a single answer and the user responding with a single set of followups, both parties iteratively build upon each other’s partial ideas. We envision that the AI’s intermediate thoughts, user feedback, clarifying questions, and real-time refinements etc form a collaborative ``thinking canvas.''

Traditional interactions are linear and unidirectional: the AI provides an answer, the user reacts, and so on. Even if the user can add further prompts, there is no persistent collaborative workspace where partial ideas accumulate and evolve.

\subsection{Implications for HCI}
The four traits of Thoughtful AI---an \textit{intermediate medium}, \textit{full-duplex process}, \textit{intrinsic driver}, and a \textit{shared cognitive space}---open up new possibilities for how humans interact with AI. Below, we discuss four implications for Human-AI Interaction.

\subsubsection{From Passive Respondents to Proactive Participants}
A first implication of Thoughtful AI is the transition from AI as a passive respondent to an active participant in interactions. Conventional AI systems operate in a reactive manner, awaiting user input before generating a response. In contrast, Thoughtful AI continuously generates thoughts, enabling it to self-initiate actions and engage more dynamically with users.

This proactivity manifests in two ways. First, the \textit{full-duplex process} (Trait 2) allows the AI to generate thoughts in parallel with user input, meaning it does not need to remain idle between interactions. It can detect emerging needs, anticipate user queries, and even interrupt when necessary. Second, thoughts equip AI with an \textit{intrinsic drive} (Trait 3) to contribute based on its internal cognitive processes. This ability to self-motivate and intervene resembles human-like initiative, distinguishing it from traditional systems that rely on heuristics or predefined triggers~\cite{horvitz1999principles}.

\subsubsection{Continuous Cognitive Alignment}

A second major implication for HCI arises from Thoughtful AI's ability to reveal \emph{in-progress} ideas (Trait 1: \textit{Intermediate Medium}) continuously (Trait 2: \textit{Full-duplex Process}). Rather than presenting a single, static final output, the system selectively shares partial thoughts. Users can then not only observe these thoughts for better interpretability, but also guide the AI's thinking early in the process, leading to more dynamic collaboration.

By showing its intermediate reasoning, Thoughtful AI helps bridge the ``Gulf of Evaluation,'' where users struggle to understand how or why a system arrived at a particular result~\cite{norman2013design}. Here, the AI’s partial thoughts provide visibility into its evolving path, allowing users to catch misconceptions or supply additional context before errors compound. Similarly, users can steer the AI's next steps more effectively, shrinking the ``Gulf of Execution,'' since they can articulate what they want \emph{as} they see the AI’s tentative directions.

This also helps build common ground~\cite{clark1991grounding} between human and AI, aligning with established theories of communication that emphasize the importance of shared context for effective collaboration. By understanding the AI's intermediate thought, users gain a clearer understanding of the system's current assumptions and partial conclusions. This shared cognitive workspace drives continuous alignment between the user's thinking and AI's thinking process.

\subsubsection{Beyond Turn-based Interaction}
Thoughtful AI prompts us to reconsider the fundamental structure of human–AI interaction. Traditional chatbot interfaces operate in discrete, back-and-forth messages, with the AI effectively going idle after sending each response. By contrast, a \emph{full-duplex process} (Trait 2) enables the AI to continuously think and listen, even when the user is not actively providing input.
This opens the door to interaction models that transcend simple chat boxes. For instance, an AI planning assistant could silently update its suggestions as it overhears new constraints in a virtual meeting, intervening only when necessary (Traits 2 and 3). 
More interestingly, AI and users can build a \emph{shared cognitive space} (Trait 4), adding thoughts, annotations, and partial ideas to a collective interface. Rather than a linear log of exchanges, this collaborative canvas allows ideas to branch, merge, and evolve continuously---resembling a dynamic mind-map or sketchnote more than a turn-based chat.

\subsubsection{Messy, Fragmented and Informal Interaction}
Traditional AI systems demand users structure their inputs as complete, well-formed queries or commands. This formality creates friction: human thinking is inherently non-linear, often involving half-formed ideas and abrupt shifts in focus. 
Thoughtful AI embraces this messiness by using intermediate thoughts as the first-class citizen of interaction. 
This shift towards a more natural, fragmented, and informal interaction style: AI thinking can be messy, incomplete, and in-progress. 
In addition, instead of requiring users to provide polished, fully-formed queries or responses, users can also input partial ideas, rough thoughts, or even keywords, and the system can process and respond to these in a similarly fragmented manner. 
This approach mirrors how humans often think and communicate: by iterating on ideas, refining them over time, and bouncing off incomplete or tentative thoughts. 
\section{Example Projects}
To further illustrate the practical applications of Thoughtful AI, we present two projects that embody its core principles. The first, Inner Thoughts, explores how AI can proactively engage in conversations by developing and evaluating its internal thoughts before participating. The second, ThinkaloudLM, investigates AI-generated thoughts as an interface component that allows users to observe and interact with in real-time.

\subsection{Inner Thoughts}
\begin{figure}
    \centering
  \includegraphics[width=0.65\linewidth]{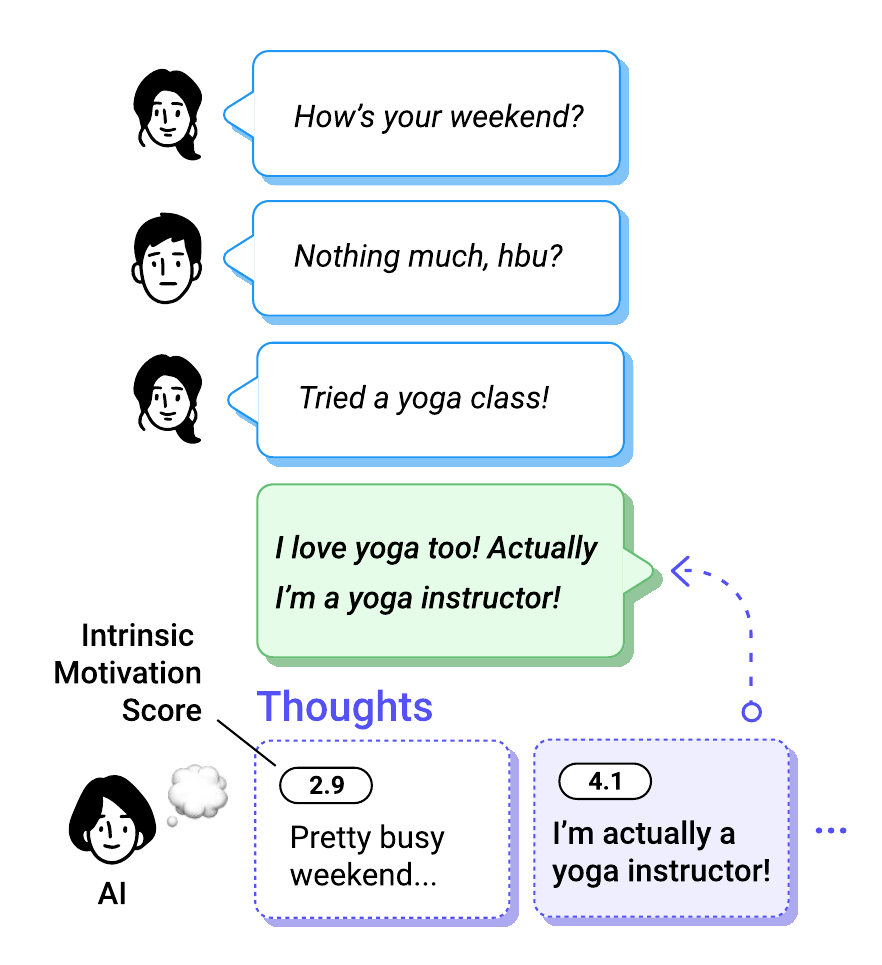}
    \caption{Conversational Agents with Inner Thoughts: AI generates a train of thoughts and evaluates them based on their intrinsic motivation to participate.}
    \label{fig:inner_thoughts}
\end{figure}

Conversational AI often rely on turn-taking prediction techniques, where an algorithm determines the most likely next speaker and generates a response accordingly. However, these approaches struggle in multi-party conversations where turn allocation is often ambiguous. Furthermore, existing conversational agents tend to fall into two extremes: either they remain passive, requiring explicit user input to respond, or they overcompensate, generating frequent and often unnecessary interruptions.

Instead of predicting conversational turns, \textit{Inner Thoughts}~\cite{liu2025inner} introduces a new method in which AI autonomously generates a continuous stream of covert (internal) thoughts, similar to how humans process conversations. These thoughts remain internal until AI evaluates whether it has sufficient \textit{intrinsic motivation} to contribute---defined by heuristics derived from a user study on human conversational behavior. Once the AI determines that a thought is relevant and meaningful, it then strategically selects an appropriate moment to engage (\autoref{fig:inner_thoughts}).

The Inner Thoughts framework embodies multiple traits of Thoughtful AI:
\textit{Intermediate Medium}: AI does not simply generate final responses but produces an evolving series of intermediate thoughts; \textit{Full-Duplex Process}: Instead of waiting for user prompts, the AI continuously listens and generates internal responses in parallel with the ongoing conversation; \textit{Intrinsic Driver}: The AI does not just react but actively determines when to contribute based on its intrinsic motivation.

The framework was implemented in two systems, a multi-agent conversational simulation and a chatbot. Through a technical evaluation, conversational agents using Inner Thoughts significantly outperformed a traditional next-speaker prediction baseline across multiple criteria, including turn appropriateness, coherence, engagement, and adaptability.
Participants overwhelmingly preferred AI interactions using Inner Thoughts (in 82\% of the conversations).

\subsection{ThinkaloudLM}
\begin{figure}
    \centering
  \includegraphics[width=\linewidth]{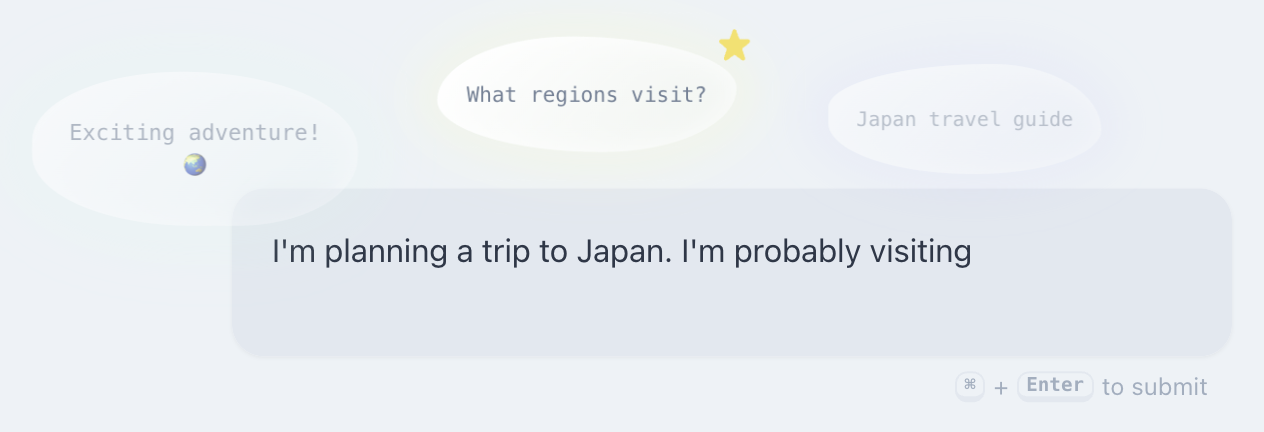}
    \caption{ThinkaloudLM: AI generates intermediate, fragmented thoughts in parellel to user input.}
    \label{fig:thinkaloudlm}
\end{figure}

While Inner Thoughts explores AI's ability to think internally, ThinkaloudLM investigates a complementary question: what if users could actively observe and engage with AI's thought process?

Most AI systems present only finalized outputs, hiding the intermediate thoughts behind their decisions. This lack of transparency can lead to user distrust, inefficiencies in communication, and missed opportunities for collaboration. For example, if an AI-powered writing assistant provides a response that feels misaligned with the user’s intent, the user must iteratively refine their query, leading to unnecessary back-and-forth exchanges.

By contrast, ThinkaloudLM envisions a system where users can engage with AI-generated ideas before they fully materialize into responses. Rather than receiving a single, static reply, users might see a live preview of AI’s thinking, including key points, possible directions, or alternative solutions. They can then refine, select, or discard thoughts before finalizing the AI’s output.

This approach leverages multiple traits of Thoughtful AI:
\textit{Intermediate Medium}: The AI’s partially formed ideas are themselves the primary interface, giving users a real-time window into the AI’s thought process; \textit{Full-duplex Process}: By continuously updating these intermediate thoughts, the AI can incorporate user feedback on-the-fly while still in the ``draft'' stage; \textit{Shared cognitive space}: Users and AI co-construct the final answer, using the AI's visible thoughts as a collaborative canvas for discussion, refinement, and mutual alignment.
To explore this concept, a participatory design study is currently underway with qualitative interviews and observational studies. 
\section{Discussion and Future Work} 
We introduced Thoughtful AI with four defining traits---\emph{intermediate medium, full-duplex process, intrinsic driver}, and \emph{shared cognitive space}---but these represent only a starting point. Below, we outline key open questions and possible directions for further exploration.

\paragraph{Are There Additional Traits of Thoughtful AI?}
Although our work has focused on these four traits, there may be additional dimensions of thoughtfulness worth exploring. For instance, what additional traits might emerge in more complex or specialized scenarios?

\paragraph{How Can We Mitigate Potential Downsides?}
An AI that continuously generates intermediate thoughts risks overwhelming users with too much information or frequent interruptions. Further, sharing internal reasoning publicly could raise privacy and security concerns. Research is needed to identify interface designs, privacy safeguards, and policies that mitigate these concerns while preserving the benefits of proactivity and collaboration.

\paragraph{What Are the Best Mechanisms for Simulating Thought?}
Current LLM-based techniques show promise, but other approaches such as symbolic reasoning or neuron-level modeling offer different trade-offs in efficiency, interpretability, and control. Future research might explore how different architectures \emph{represent} and \emph{execute} continuous thought.

\paragraph{How Can Thoughtfulness Be Embedded in AI Architecture and Training?}
Related to the simulation question is how to integrate thoughtful behavior at the architectural and training levels. Current practices often rely on prompting an off-the-shelf LLM for reasoning traces (e.g., chain-of-thought prompting). However, more direct ways of embedding thoughtfulness could be promising.

\paragraph{How Might Thoughtful AI Influence Human Thinking?}
If AI continuously generates and shares evolving thoughts, how might this reshape human cognitive processes? Will users offload more cognitive tasks to AI? Alternatively, could exposure to AI's thoughts improve human metacognition by encouraging users to articulate and refine their own thought processes? Additionally, how might this affect creative workflows, decision-making, and critical thinking skills?

\paragraph{What New Interaction Paradigms Can Emerge from Thoughtful AI?}
Finally, Thoughtful AI has the potential to reshape how users and AI collaborate, moving beyond simple chat interfaces. Exploring these new paradigms will be crucial to realizing AI that \emph{thinks with us}, rather than simply \emph{responding to us}.

\section{Conclusion}
In this position paper, we proposed the concept of Thoughtful AI, a new paradigm in which AI continuously generates, develops, and selectively communicates its evolving thoughts throughout an interaction. By moving beyond the conventional turn-based, input-output framework, Thoughtful AI enables more proactive, adaptive, and collaborative engagement between humans and AI. We identified four key traits, \textit{intermediate medium, full-duplex process, intrinsic driver}, and \textit{shared cognitive space}, and demonstrated their potential through two example projects.

However, Thoughtful AI is not merely about enhancing existing AI assistants; it represents a broader shift toward more \textit{fluid, intuitive, and adaptive} paradigms of human–AI interaction. At its core, this approach challenges the decades-old model of AI as a passive respondent confined to chat interfaces---a structure that has persisted since the era of ELIZA~\cite{Weizenbaum1966ELIZAaCP}. 
We advocate for a fundamental reimagining of AI systems and human-AI interaction, ones that enables richer collaboration, fosters deeper trust, and ultimately creates a more natural way for humans and machines to think together.

\bibliographystyle{ACM-Reference-Format}
\bibliography{refs}


\begin{thebibliography}{14}


\ifx \showCODEN    \undefined \def \showCODEN     #1{\unskip}     \fi
\ifx \showISBNx    \undefined \def \showISBNx     #1{\unskip}     \fi
\ifx \showISBNxiii \undefined \def \showISBNxiii  #1{\unskip}     \fi
\ifx \showISSN     \undefined \def \showISSN      #1{\unskip}     \fi
\ifx \showLCCN     \undefined \def \showLCCN      #1{\unskip}     \fi
\ifx \shownote     \undefined \def \shownote      #1{#1}          \fi
\ifx \showarticletitle \undefined \def \showarticletitle #1{#1}   \fi
\ifx \showURL      \undefined \def \showURL       {\relax}        \fi
\providecommand\bibfield[2]{#2}
\providecommand\bibinfo[2]{#2}
\providecommand\natexlab[1]{#1}
\providecommand\showeprint[2][]{arXiv:#2}

\bibitem[Clark and Brennan(1991)]%
        {clark1991grounding}
\bibfield{author}{\bibinfo{person}{Herbert~H. Clark} {and} \bibinfo{person}{Susan~E. Brennan}.} \bibinfo{year}{1991}\natexlab{}.
\newblock \showarticletitle{Grounding in Communication}.
\newblock In \bibinfo{booktitle}{\emph{Perspectives on Socially Shared Cognition}}, \bibfield{editor}{\bibinfo{person}{Lauren~B. Resnick}, \bibinfo{person}{John~M. Levine}, {and} \bibinfo{person}{Stephanie~D. Teasley}} (Eds.). \bibinfo{publisher}{American Psychological Association}, \bibinfo{pages}{127--149}.
\newblock


\bibitem[Descartes(1996)]%
        {descartes1996discourse}
\bibfield{author}{\bibinfo{person}{Ren{\'e} Descartes}.} \bibinfo{year}{1996}\natexlab{}.
\newblock \bibinfo{booktitle}{\emph{Discourse on the method: And, meditations on first philosophy}}.
\newblock \bibinfo{publisher}{Yale University Press}.
\newblock


\bibitem[{Google DeepMind}(2025)]%
        {deepmind2025geminiflashthinking}
\bibfield{author}{\bibinfo{person}{{Google DeepMind}}.} \bibinfo{year}{2025}\natexlab{}.
\newblock \bibinfo{title}{Gemini Flash Thinking}.
\newblock
\urldef\tempurl%
\url{https://deepmind.google/technologies/gemini/flash-thinking/}
\showURL{%
\tempurl}
\newblock
\shownote{Accessed: 2025-02-13}.


\bibitem[Guo et~al\mbox{.}(2025)]%
        {guo2025deepseek}
\bibfield{author}{\bibinfo{person}{Daya Guo}, \bibinfo{person}{Dejian Yang}, \bibinfo{person}{Haowei Zhang}, \bibinfo{person}{Junxiao Song}, \bibinfo{person}{Ruoyu Zhang}, \bibinfo{person}{Runxin Xu}, \bibinfo{person}{Qihao Zhu}, \bibinfo{person}{Shirong Ma}, \bibinfo{person}{Peiyi Wang}, \bibinfo{person}{Xiao Bi}, {et~al\mbox{.}}} \bibinfo{year}{2025}\natexlab{}.
\newblock \showarticletitle{Deepseek-r1: Incentivizing reasoning capability in llms via reinforcement learning}.
\newblock \bibinfo{journal}{\emph{arXiv preprint arXiv:2501.12948}} (\bibinfo{year}{2025}).
\newblock


\bibitem[Horvitz(1999)]%
        {horvitz1999principles}
\bibfield{author}{\bibinfo{person}{Eric Horvitz}.} \bibinfo{year}{1999}\natexlab{}.
\newblock \showarticletitle{Principles of mixed-initiative user interfaces}. In \bibinfo{booktitle}{\emph{Proceedings of the SIGCHI Conference on Human Factors in Computing Systems}} (Pittsburgh, Pennsylvania, USA) \emph{(\bibinfo{series}{CHI '99})}. \bibinfo{publisher}{Association for Computing Machinery}, \bibinfo{address}{New York, NY, USA}, \bibinfo{pages}{159–166}.
\newblock
\showISBNx{0201485591}
\href{https://doi.org/10.1145/302979.303030}{doi:\nolinkurl{10.1145/302979.303030}}


\bibitem[Jaech et~al\mbox{.}(2024)]%
        {jaech2024openai}
\bibfield{author}{\bibinfo{person}{Aaron Jaech}, \bibinfo{person}{Adam Kalai}, \bibinfo{person}{Adam Lerer}, \bibinfo{person}{Adam Richardson}, \bibinfo{person}{Ahmed El-Kishky}, \bibinfo{person}{Aiden Low}, \bibinfo{person}{Alec Helyar}, \bibinfo{person}{Aleksander Madry}, \bibinfo{person}{Alex Beutel}, \bibinfo{person}{Alex Carney}, {et~al\mbox{.}}} \bibinfo{year}{2024}\natexlab{}.
\newblock \showarticletitle{Openai o1 system card}.
\newblock \bibinfo{journal}{\emph{arXiv preprint arXiv:2412.16720}} (\bibinfo{year}{2024}).
\newblock


\bibitem[Liu et~al\mbox{.}(2025)]%
        {liu2025inner}
\bibfield{author}{\bibinfo{person}{Xingyu~Bruce Liu}, \bibinfo{person}{Shitao Fang}, \bibinfo{person}{Weiyan Shi}, \bibinfo{person}{Chien-Sheng Wu}, \bibinfo{person}{Takeo Igarashi}, {and} \bibinfo{person}{Xiang~Anthony Chen}.} \bibinfo{year}{2025}\natexlab{}.
\newblock \showarticletitle{Proactive Conversational Agents with Inner Thoughts}. In \bibinfo{booktitle}{\emph{Proceedings of the 2025 CHI Conference on Human Factors in Computing Systems}} (Yokohama, Japan) \emph{(\bibinfo{series}{CHI '25})}. \bibinfo{publisher}{Association for Computing Machinery}, \bibinfo{address}{New York, NY, USA}.
\newblock
\href{https://doi.org/10.1145/3706598.3713760}{doi:\nolinkurl{10.1145/3706598.3713760}}


\bibitem[Miller and Cohen(2001)]%
        {miller2001integrative}
\bibfield{author}{\bibinfo{person}{Earl~K Miller} {and} \bibinfo{person}{Jonathan~D Cohen}.} \bibinfo{year}{2001}\natexlab{}.
\newblock \showarticletitle{An integrative theory of prefrontal cortex function}.
\newblock \bibinfo{journal}{\emph{Annual review of neuroscience}} \bibinfo{volume}{24}, \bibinfo{number}{1} (\bibinfo{year}{2001}), \bibinfo{pages}{167--202}.
\newblock


\bibitem[Newell(1972)]%
        {newell1972human}
\bibfield{author}{\bibinfo{person}{Allen Newell}.} \bibinfo{year}{1972}\natexlab{}.
\newblock \showarticletitle{Human problem solving}.
\newblock \bibinfo{journal}{\emph{Upper Saddle River/Prentive Hall}} (\bibinfo{year}{1972}).
\newblock


\bibitem[Norman(2013)]%
        {norman2013design}
\bibfield{author}{\bibinfo{person}{Donald~A. Norman}.} \bibinfo{year}{2013}\natexlab{}.
\newblock \bibinfo{booktitle}{\emph{The Design of Everyday Things} (\bibinfo{edition}{revised and expanded} ed.)}.
\newblock \bibinfo{publisher}{Basic Books}.
\newblock
\showISBNx{978-0465050659}


\bibitem[Raichle et~al\mbox{.}(2001)]%
        {raichle2001default}
\bibfield{author}{\bibinfo{person}{Marcus~E Raichle}, \bibinfo{person}{Ann~Mary MacLeod}, \bibinfo{person}{Abraham~Z Snyder}, \bibinfo{person}{William~J Powers}, \bibinfo{person}{Debra~A Gusnard}, {and} \bibinfo{person}{Gordon~L Shulman}.} \bibinfo{year}{2001}\natexlab{}.
\newblock \showarticletitle{A default mode of brain function}.
\newblock \bibinfo{journal}{\emph{Proceedings of the national academy of sciences}} \bibinfo{volume}{98}, \bibinfo{number}{2} (\bibinfo{year}{2001}), \bibinfo{pages}{676--682}.
\newblock


\bibitem[Turing(1950)]%
        {Turing1950ComputingMA}
\bibfield{author}{\bibinfo{person}{Alan~M. Turing}.} \bibinfo{year}{1950}\natexlab{}.
\newblock \showarticletitle{Computing Machinery and Intelligence}.
\newblock \bibinfo{journal}{\emph{Mind}}  \bibinfo{volume}{LIX} (\bibinfo{year}{1950}), \bibinfo{pages}{433--460}.
\newblock
\urldef\tempurl%
\url{https://api.semanticscholar.org/CorpusID:14636783}
\showURL{%
\tempurl}


\bibitem[Wei et~al\mbox{.}(2022)]%
        {wei2022chain}
\bibfield{author}{\bibinfo{person}{Jason Wei}, \bibinfo{person}{Xuezhi Wang}, \bibinfo{person}{Dale Schuurmans}, \bibinfo{person}{Maarten Bosma}, \bibinfo{person}{Fei Xia}, \bibinfo{person}{Ed Chi}, \bibinfo{person}{Quoc~V Le}, \bibinfo{person}{Denny Zhou}, {et~al\mbox{.}}} \bibinfo{year}{2022}\natexlab{}.
\newblock \showarticletitle{Chain-of-thought prompting elicits reasoning in large language models}.
\newblock \bibinfo{journal}{\emph{Advances in neural information processing systems}}  \bibinfo{volume}{35} (\bibinfo{year}{2022}), \bibinfo{pages}{24824--24837}.
\newblock


\bibitem[Weizenbaum(1966)]%
        {Weizenbaum1966ELIZAaCP}
\bibfield{author}{\bibinfo{person}{Joseph Weizenbaum}.} \bibinfo{year}{1966}\natexlab{}.
\newblock \showarticletitle{ELIZA—a computer program for the study of natural language communication between man and machine}.
\newblock \bibinfo{journal}{\emph{Commun. ACM}}  \bibinfo{volume}{9} (\bibinfo{year}{1966}), \bibinfo{pages}{36 -- 45}.
\newblock
\urldef\tempurl%
\url{https://api.semanticscholar.org/CorpusID:1896290}
\showURL{%
\tempurl}


\end{thebibliography}

\end{document}